\documentclass{PoS}
\usepackage{amsmath}
\usepackage{txfonts}
\usepackage{subfigure}
\usepackage{ulem}

\newcommand{\etal}{{\it et al.}}

\newcommand{\cf}[1]{{Fig.~\ref{#1}}}

\newcommand{\Br}{{\rm Br}}

\newcommand{\beq}[1]{
\begin{equation}\label{#1}}
\newcommand{\eeq}{\end{equation}}
\newcommand{\bea}[1]{
\begin{eqnarray}\label{#1}}
\newcommand{\eea}{\end{eqnarray}}
\usepackage{lineno}

\newcommand{\out}{\raise-3pt\hbox{\scriptsize    out}}

\title{Total $J/\psi$ and $\Upsilon$ production cross section at the LHC: theory vs. experiment}

\ShortTitle{Total $J/\psi$ and $\Upsilon$ production cross section at the LHC}

\author{J.P. Lansberg\\
       IPNO, Universit\'e Paris-Sud 11, CNRS/IN2P3, F-91406, Orsay, France\thanks{Permanent address at IPNO} \\
Centre de Physique Th\'eorique, \'Ecole polytechnique, CNRS, F-91128, Palaiseau, France\\
E-mail: \email{Jean-Philippe.Lansberg@in2p3.fr}}

\abstract{We evaluate the production cross section for direct $J/\psi$ and $\Upsilon$
integrated in $P_T$ for various collision energies  in the QCD-based Colour-Singlet Model (CSM).
We consider the LO contribution from gluon fusion whose $P_T$-integrated cross section 
shows a very good agreement with the Tevatron and LHC data, both for 
$J/\psi$ and $\Upsilon$. The rapidity distribution of this yield
is evaluated in the central region relevant for the ATLAS and CMS detectors, as well as in the 
more forward region relevant for the ALICE and LHCb detectors. The results obtained here 
are compatible with those of other approaches within the range of the theoretical uncertainties
which are admittedly very large. This suggests that the ``mere'' measurements of the yield at the LHC will not help 
disentangle between the different possible quarkonium production mechanisms. Yet, the comparison
with the first LHC results by ALICE, ATLAS, CMS and LHCb confirms that the CSM correctly accounts for the
$P_T$-integrated yield at $\sqrt{s}=7$ TeV.}

\FullConference{35th International Conference of High Energy Physics - ICHEP2010,\\
		July 22-28, 2010\\
		Paris France}

\begin{document}

\section{Introduction}

In 2007, the first evaluations of QCD corrections to quarkonium-production rates at hadron colliders 
became available. It is now widely accepted that $\alpha^4_s$ and $\alpha^5_s$ 
corrections to the CSM~\cite{CSM_hadron} are fundamental for understanding the $P_T$ spectrum of 
$J/\psi$ and $\Upsilon$ produced in high-energy hadron  
collisions~\cite{Campbell:2007ws,Artoisenet:2007xi,Gong:2008sn,Artoisenet:2008fc,Artoisenet:2008zza,Lansberg:2008gk},
while the difficulties of predicting these observables had been initially attributed to 
non-perturbative effects associated with channels in which the heavy
quark and antiquark are produced in a colour-octet 
state~\cite{Lansberg:2006dh,Brambilla:2004wf,Kramer:2001hh,Lansberg:2008zm}.
Further, the effect of QCD corrections is also manifest in the polarisation predictions. While the 
$J/\psi$ and $\Upsilon$ produced inclusively or in association with a 
photon are predicted to be transversally polarised at LO, it has been recently emphasised that their 
polarisation at NLO is  increasingly longitudinal when $P_T$ gets 
larger~\cite{Gong:2008sn,Artoisenet:2008fc,Li:2008ym,Lansberg:2009db,Lansberg:2010vq}.

In a recent work~\cite{Brodsky:2009cf}, we have also shown that hard subprocesses based on colour 
singlet $Q \bar Q$ configurations alone are sufficient to account for the observed magnitude of 
the $P_T$-integrated cross section. In particular, the predictions for the $J/\psi$ yield at LO~\cite{CSM_hadron} 
 and NLO~\cite{Campbell:2007ws,Artoisenet:2007xi,Gong:2008sn} 
accuracy are both compatible with the measurements by the PHENIX collaboration at 
RHIC~\cite{Adare:2006kf} within the present uncertainties. This also pointed at a reduced impact
of the $s$-channel cut contributions~\cite{Haberzettl:2007kj} as well as of some specific colour-octet mediated 
channels relevant for the low $P_T$ region ($^1S_0^{[8]}$ and  $^3P_J^{[8]}$). The latter are anyway very strongly constrained
by very important recent $e^+e^-$ analyses~\cite{ee} which leave in some cases no room at all for colour octets of any kind.

 The compatibility between the LO and NLO yields provided  some 
indications that the computations are carried in a proper perturbative regime, at least at RHIC energies.
The agreement with the data is improved when hard subprocesses involving the 
charm-quark distribution of the colliding protons are taken into consideration. These 
constitute part of the LO ($\alpha_S^3$) rate  and can be
responsible for a significant fraction of the observed yield~\cite{Brodsky:2009cf,Lansberg:2010kh}.

We proceed here to the evaluation the $P_T$-integrated yield at higher energies both 
in the central and forward rapidity regions. At large energies, our study shows that 
the theoretical uncertainties on the $J/\psi$ yield for become very large --close to one 
decade-- reminiscent of the case of total charm production~\cite{Vogt:1900zz}. 
Both for the $J/\psi$ and the $\Upsilon$, we find a very good agreement with the CDF measurements~\cite{Acosta:2004yw} and 
the first LHC ones. Finally, we shortly discuss the impact of higher QCD corrections 
and the comparison with other approaches.

\section{Total $J/\psi$ and $\Upsilon$ cross section at the LHC\protect\footnote{Note that we have not depicted the systematic
uncertainties attached the unknown $J/\psi$ and $\Upsilon$ polarisation produced at the LHC . They can be as high as 50 $\%$ at very low $P_T$ for
extreme configurations.}}

The $P_T$ integrated cross sections obtained here have been evaluated along the 
same lines as our previous study~\cite{Brodsky:2009cf}. The uncertainty bands have been evaluated 
following exactly the same procedure using the same values for $m_c$ ($m_b$), $\mu_R$ and $\mu_F$.

\begin{figure}[hbt!]
  \begin{center} 
    \begin{minipage}[b]{0.48\textwidth}%
      \centerline{
    \includegraphics[width=\textwidth,clip=true]{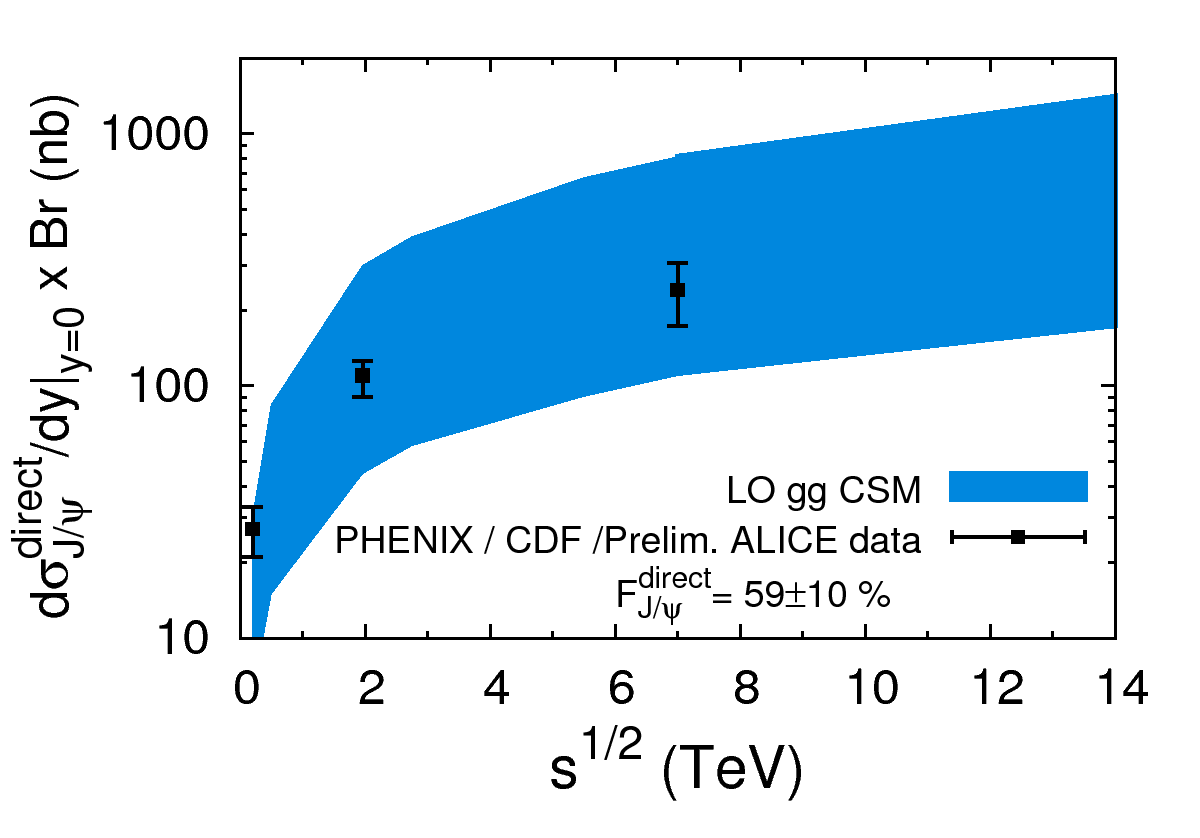}  
      }\vspace*{-.5cm}
      \caption{
$d\sigma^{direct}_{J/\psi}/dy|_{y=0}\times \Br$ from $gg$ fusion in $pp$ collisions for $\sqrt{s}$ from 200 GeV up to 14 TeV
compared to the PHENIX~\protect\cite{Adare:2006kf}, CDF~\protect\cite{Acosta:2004yw} and ALICE~\protect\cite{boyer} data  multiplied by the direct fraction (see text).\protect\\
\label{sigma_vs_s-jpsi}
      }
    \end{minipage} 
    \hfill
    \begin{minipage}[b]{0.48\textwidth}%
      \centerline{
   \includegraphics[width=\textwidth,clip=true]{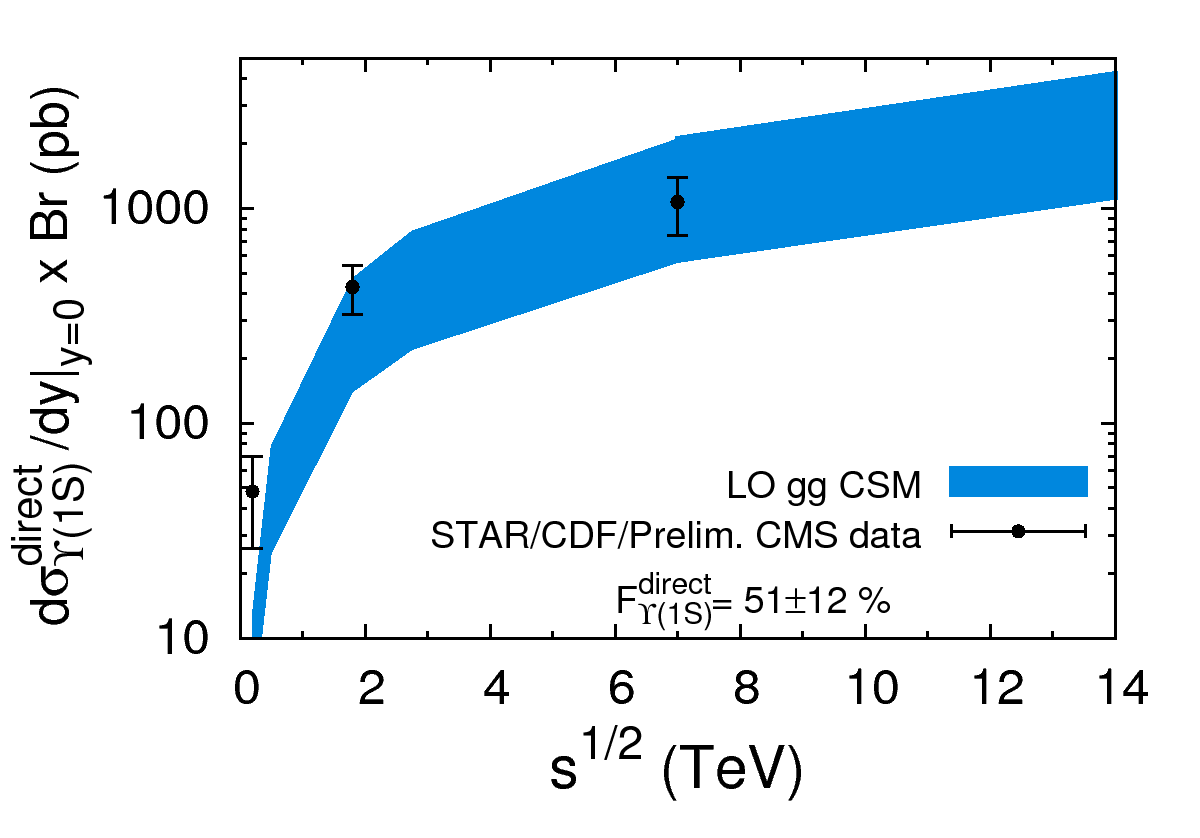}  
      }\vspace*{-.5cm}
\caption{$d\sigma^{direct}_{\Upsilon}/dy|_{y=0}\times \Br$ from $gg$ fusion in $pp$ collisions for $\sqrt{s}$ from 200 GeV up to 14 TeV
compared to the STAR~\protect\cite{Abelev:2010am}, CDF~\protect\cite{Acosta:2001gv} and CMS~\protect\cite{CMS-Upsilon}
data  multiplied by the direct fraction (see text).\protect\\
}
\label{sigma_vs_s-upsilon}
      \end{minipage}\vspace*{-1cm}
    \end{center}
\end{figure}

In \cf{sigma_vs_s-jpsi}, we show $d\sigma^{direct}_{J/\psi}/dy|_{y=0}\times \Br$  as function
of  $\sqrt{s}$ from 200 GeV up to 14 TeV compared to the PHENIX~\protect\cite{Adare:2006kf}, the CDF~\protect\cite{Acosta:2004yw} and ALICE~\protect\cite{boyer} data  multiplied by the direct fraction\footnote{We employ this makeshift, although the $\chi_c$ yield can now be  computed at NLO accuracy~\cite{Ma:2010vd}, since one cannot extend these computations to low $P_T$.}\footnote{Note that the measurement 
of the prompt yield  by CDF went only down to $P_T=1.25$ GeV. 
We have assumed a fraction of non-prompt $J/\psi$ of $10 \%$ below. We have assumed the same fraction at $\sqrt{s}=7$ TeV for the ALICE 
data.} measured by CDF at $\sqrt{s}=1.8$ TeV~\cite{Abe:1997yz}. We have found  a good agreement. At larger energies, these results at 7 TeV (100 to 800 nb) and at 14 TeV (200 to 
1400 nb) are in the same range as those of the Colour Evaporation Model~\cite{Bedjidian:2004gd} 
with central (upper) values of 140 nb (400 nb) at 7 TeV  and  200 nb (550 nb) at 14 TeV. 
They are also compatible with the results of the ''gluon tower model'' (GTM)~\cite{Khoze:2004eu}, 
300  nb at  7 TeV  and 480 nb at 14 TeV,  
which takes into account some NNLO contributions  shown to be enhanced by $\log(s)$. 
Quoting the authors~\cite{Khoze:2004eu},  ``the expected accuracy of the prediction is about a factor of 2-3 in either 
direction or even worse.'' 
In \cf{sigma_vs_s-upsilon}, we show the same predictions  for direct $\Upsilon$
for $\sqrt{s}$ from 200 GeV up to 14 TeV compared to the STAR~\protect\cite{Abelev:2010am}, 
CDF~\protect\cite{Acosta:2001gv} and CMS~\protect\cite{CMS-Upsilon}
data  multiplied by the direct fraction measured by CDF~\cite{Affolder:1999wm} at $\sqrt{s}=1.8$ TeV.
While there seems to be some tension between the CSM predictions and the data at $\sqrt{s}=200$ GeV, 
the agreement is very good with the preliminary CMS data.

\begin{figure}[hbt!]
  \begin{center} 
    \begin{minipage}[b]{0.48\textwidth}%
      \centerline{
 \includegraphics[width=\textwidth,clip=true]{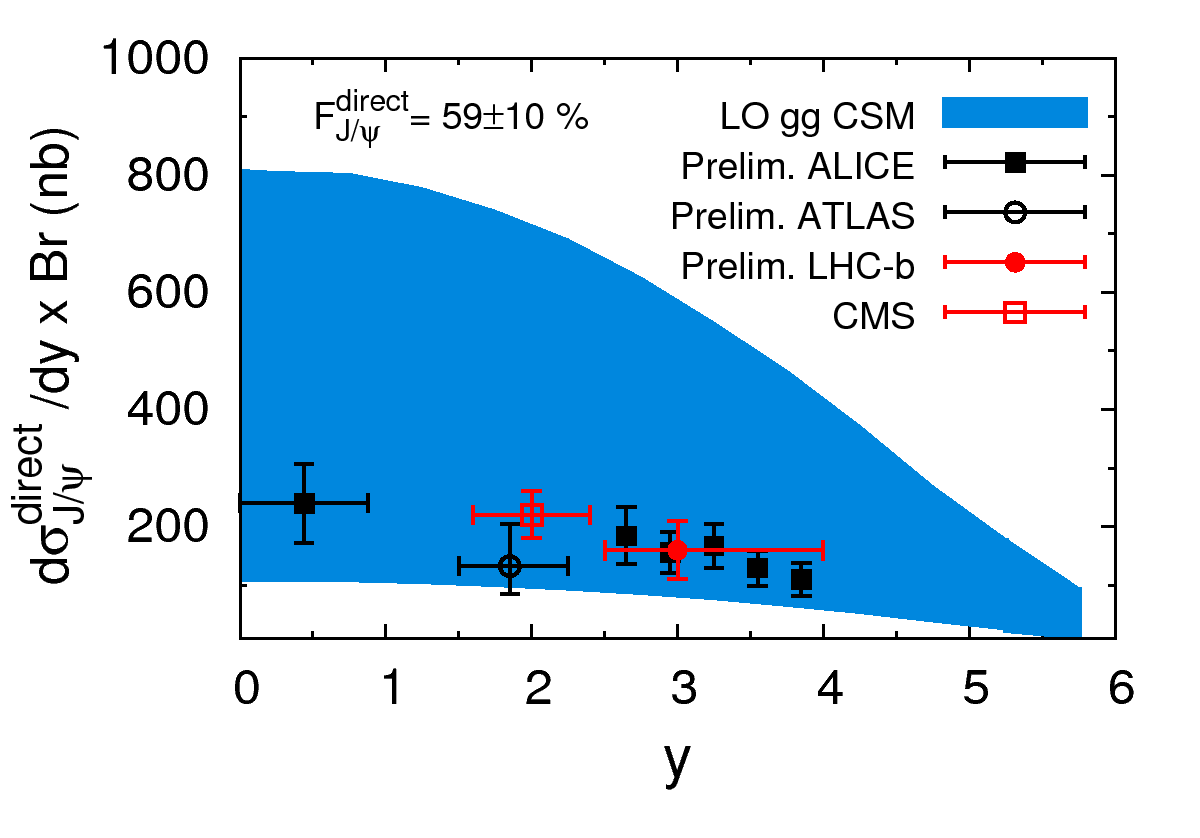}
      }\vspace*{-.5cm}
      \caption{$d\sigma^{direct}_{J/\psi}/dy\times \Br$ from $gg$ fusion LO contributions in $pp$ collisions at $\sqrt{s}=7$ TeV compared
to ALICE~\protect\cite{boyer}, ATLAS~\protect\cite{ATLAS-JPsi}, CMS~\protect\cite{Collaboration:2010yr}
and LHCb~\protect\cite{LHCb-JPsi}  results multiplied by the direct fraction (and the prompt fraction (10 $\%$) if applicable). 
\label{7tev-Jpsi}
      }
    \end{minipage} 
    \hfill
    \begin{minipage}[b]{0.5\textwidth}%
      \centerline{
   \includegraphics[width=\textwidth,clip=true]{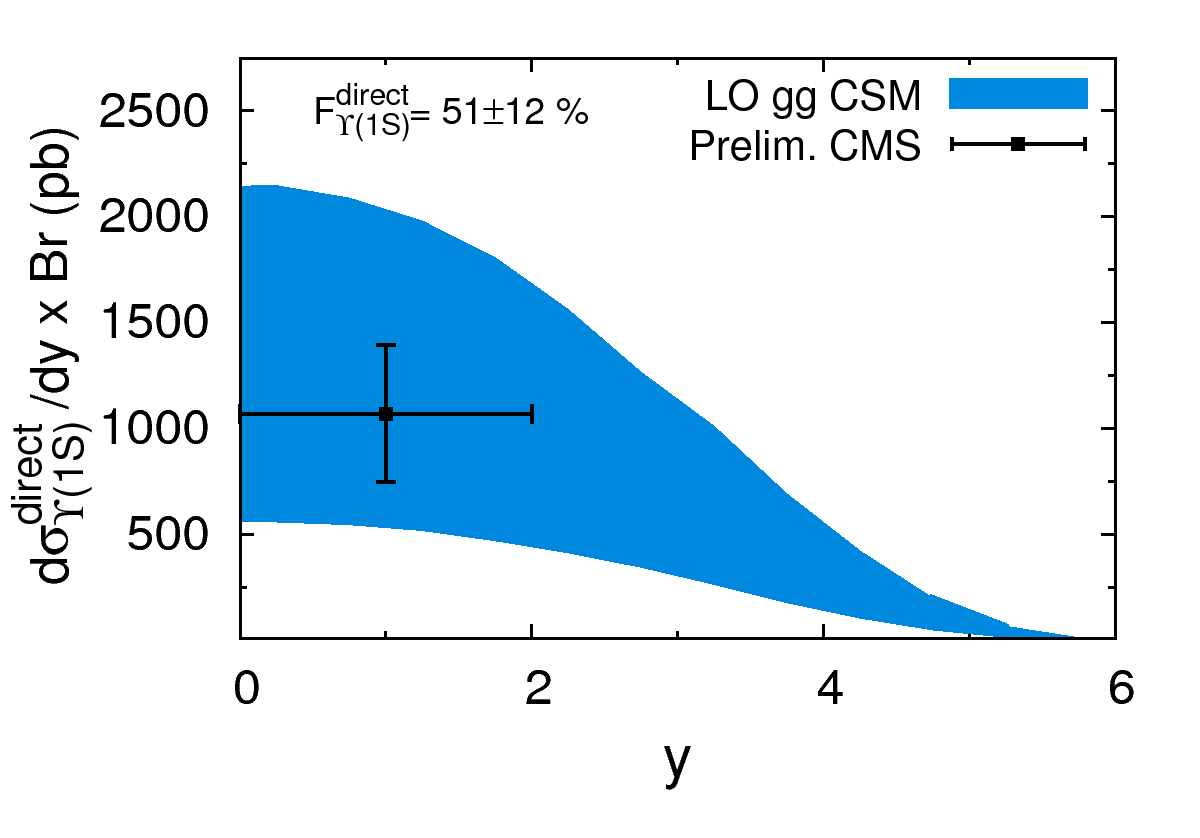}  
      }\vspace*{-.5cm}
\caption{$d\sigma^{direct}_{\Upsilon}/dy\times \Br$ from LO CSM via $gg$ fusion in $pp$ collisions at $\sqrt{s}=7$ TeV
compared to the preliminary CMS results~\protect\cite{CMS-Upsilon} multiplied by the direct fraction (see text). \protect\\ }
\label{7tev-Upsi}
      \end{minipage}\vspace*{-.5cm}
    \end{center}
\end{figure}

In \cf{7tev-Jpsi}, we show the  cross section differential in rapidity at $\sqrt{s}=7$ TeV compared to the 
ALICE~\protect\cite{boyer}, ATLAS~\protect\cite{ATLAS-JPsi}, CMS~\protect\cite{Collaboration:2010yr} and  LHCb~\protect\cite{LHCb-JPsi} 
 results multiplied by the direct fraction (and the prompt fraction (10 $\%$) if applicable)
as done for~\cf{sigma_vs_s-jpsi}. The region $y<0$ is not plotted since it does not contain any additional physical information. 
Similarly, experimental measurements for $y>0$ or $y<0$ are physically strictly equivalent. In \cf{7tev-Upsi}, we show the same plot
for the direct $\Upsilon$ yield along the only public results available so far, from CMS~\cite{CMS-Upsilon}, multiplied
by the direct fraction from CDF. In both cases, the agreement is found to be very good.

\section{Discussion and conclusion}

Let us now briefly discuss  the expectations for the results when QCD corrections are taken into account. 
First, we would like to stress that, although NLO results~\cite{Campbell:2007ws} are perfectly well behaved 
in nearly all of the phase-space region at RHIC energies~\cite{Brodsky:2009cf}, it does not seem to be 
so for larger $s$ for the $J/\psi$. One observes that the region where the differential cross section in $P_T$ and/or $y$ 
is negative (i.e. very low $P_T$ and large $y$) widens for increasing $s$. Negative differential cross 
sections at low $P_T$ are a known issue. Nonetheless, for $\sqrt{s}$ above a couple of TeV, and for some (common) choices of 
$\mu_F$ and $\mu_R$, the $P_T$-integrated ``yield'' happens to become negative, even in the central region. 
This can of course be explained by a larger contribution from the virtual corrections at $\alpha_S^4$ 
--which can be negative-- compared to the real emission contributions --which are positive--. Naturally, 
such results cannot be compared to experimental ones. 
This also points at virtual NNLO 
contributions at low $P_T$ which are likely large but which are not presently known. Yet, as already mentioned, 
specific NNLO contributions were shown~\cite{Khoze:2004eu} to be enhanced by $\log(s)$. We also note that 
this issue
of negative cross sections at LHC energies  does not seem as critical for the
$\Upsilon$, whose production is initiated by gluons with larger $x_{Bj}$.

As we have  discussed above, one may try to compare the LO CSM 
with other theoretical approaches such as the CEM~\cite{Bedjidian:2004gd} and the GTM~\cite{Khoze:2004eu}. 
They all qualitatively agree, as well as with existing measurements. 
For all approaches, one expects a significant spread --up to a factor of ten~-- 
of the results when the scales and the mass are varied. 

Owing to these uncertainties, it will be difficult to discriminate between different 
mechanisms~\cite{Lansberg:2006dh,Brambilla:2010cs} by only 
relying on the yield integrated in $P_T$ and even, to a less extent, on its $P_T$ dependent counterpart. 
This is a clear motivation to study at the LHC other observables related to the production of $J/\psi$ 
such as its production in association with a single charm (or lepton)~\cite{Brodsky:2009cf}, with a prompt isolated 
photon~\cite{Li:2008ym,Lansberg:2009db} or even with a pair of $c \bar c$~\cite{Artoisenet:2007xi}. 

Yet, the good agreement at the level of the $P_T$-integrated cross section between the CSM and the available measurements
from $\sqrt{s}=200$ GeV to $\sqrt{s}=7$ TeV  confirms the little --if not negligible-- impact of colour-octet contributions
at low $P_T$, at least for the $J/\psi$,  in accordance to recent  $e^+e^-$ analyses~\cite{ee}.

\section*{Acknowledgments}

I thank V. Khoze, A. Kraan, M. Ryskin, E. Scomparin, G. Smbat, R. Vogt for correspondences,
B. Boyer, F. Fleuret, J. He, G. Martinez, P. Robbe and C. Suire for useful discussions.


\begin{thebibliography}{99}


\bibitem{CSM_hadron}
C-H. Chang,
{Nucl. Phys. } B {\bf 172} (1980) 425; 
R. Baier and R. R\"uckl,
{Phys. Lett. } B {\bf 102} (1981) 364;
  E.~L.~Berger and D.~L.~Jones,
  Phys.\ Rev.\  D {\bf 23} (1981) 1521;
R. Baier and R. R\"uckl,
{Z. Phys. } C {\bf 19} (1983) 251;
  V.~G.~Kartvelishvili, A.~K.~Likhoded and S.~R.~Slabospitsky,
  Sov.\ J.\ Nucl.\ Phys.\  {\bf 28} (1978) 678
  [Yad.\ Fiz.\  {\bf 28} (1978) 1315].





%
\vspace*{-0.2cm}\bibitem{Campbell:2007ws}
  J.~Campbell, F.~Maltoni and F.~Tramontano,
  Phys.\ Rev.\ Lett.\  {\bf 98}, 252002 (2007)



\vspace*{-0.2cm}\bibitem{Artoisenet:2007xi}
  P.~Artoisenet, J.~P.~Lansberg and F.~Maltoni,
  Phys.\ Lett.\  B {\bf 653}, 60 (2007)




\vspace*{-0.2cm}\bibitem{Gong:2008sn}
  B.~Gong and J.~X.~Wang,
  Phys.\ Rev.\ Lett.\  {\bf 100} (2008) 232001; 
  Phys.\ Rev.\  D {\bf 78} (2008) 074011.



\vspace*{-0.2cm}\bibitem{Artoisenet:2008fc}
  P.~Artoisenet, J.~Campbell, J.~P.~Lansberg, F.~Maltoni and F.~Tramontano,
  Phys.\ Rev.\ Lett.\  {\bf 101} (2008) 152001.

\vspace*{-0.2cm}\bibitem{Artoisenet:2008zza}
  P.~Artoisenet,
  AIP Conf.\ Proc.\  {\bf 1038}, 55 (2008);

\vspace*{-0.2cm}\bibitem{Lansberg:2008gk}
  J.~P.~Lansberg,
  Eur.\ Phys.\ J.\  C {\bf 61}, 693 (2009);




\vspace*{-0.2cm}\bibitem{Lansberg:2006dh}
  J.~P.~Lansberg,
  Int.\ J.\ Mod.\ Phys.\  A {\bf 21}, 3857 (2006)

\vspace*{-0.2cm}\bibitem{Brambilla:2004wf}
  N.Brambilla {\it et al.}, CERN Yellow Report 2005-005,
hep-ph/0412158

\vspace*{-0.2cm}\bibitem{Kramer:2001hh}
  M.~Kramer,
  Prog.\ Part.\ Nucl.\ Phys.\  {\bf 47}, 141 (2001)

\vspace*{-0.2cm}\bibitem{Lansberg:2008zm}
  J.~P.~Lansberg {\it et al.},
  AIP Conf.\ Proc.\  {\bf 1038} (2008) 15
  [arXiv:0807.3666 [hep-ph]].





\vspace*{-0.2cm}\bibitem{Li:2008ym}
  R.~Li and J.~X.~Wang,
  Phys.\ Lett.\  B {\bf 672} (2009) 51.

\vspace*{-0.2cm}\bibitem{Lansberg:2009db}
  J.~P.~Lansberg,
  Phys.\ Lett.\  B {\bf 679} (2009) 340.


\vspace*{-0.2cm}\bibitem{Lansberg:2010vq}
  J.~P.~Lansberg,
  Phys.\ Lett.\  B {\bf 695} (2011) 149.


\vspace*{-0.2cm}\bibitem{Brodsky:2009cf}
  S.~J.~Brodsky and J.~P.~Lansberg,
  Phys.\ Rev.\ D {\bf 81} 051502(R) (2010).



\vspace*{-0.2cm}\bibitem{Adare:2006kf}
  A.~Adare {\it et al.}, 
  Phys.\ Rev.\ Lett.\  {\bf 98} (2007) 232002.
  C.~L.~da Silva,
 Nucl.\ Phys.\  A {\bf 830} (2009) 227C; 
  L.~Linden~Levy,
Nucl.\ Phys.\  A {\bf 830} (2009) 353C




\vspace*{-0.2cm}\bibitem{Haberzettl:2007kj}
  H.~Haberzettl and J.~P.~Lansberg,
  Phys.\ Rev.\ Lett.\  {\bf 100} (2008) 032006.
  J.~P.~Lansberg, J.~R.~Cudell and Yu.~L.~Kalinovsky,
  Phys.\ Lett.\ B {\bf 633} (2006) 301.




\vspace*{-0.2cm}\bibitem{ee}
  Z.~G.~He, Y.~Fan and K.~T.~Chao,
  Phys.\ Rev.\  D {\bf 81} (2010) 054036;
  Y.~J.~Zhang, Y.~Q.~Ma, K.~Wang and K.~T.~Chao,
  Phys.\ Rev.\  D {\bf 81} (2010) 034015;
Y.~Q.~Ma, Y.~J.~Zhang and K.~T.~Chao,
Phys.\ Rev.\ Lett.\ {\bf 102} (2009) 162002;
B.~Gong and J.~X.~Wang,
Phys.\ Rev.\ Lett.\ {\bf 102} (2009) 162003.



\vspace*{-0.2cm}\bibitem{Vogt:1900zz}
  R.~Vogt,
  Eur.\ Phys.\ J.\  C {\bf 61} (2009) 793.

\vspace*{-0.2cm}\bibitem{Lansberg:2010kh}
  J.~P.~Lansberg,
  to appear in {\it La Thuile 2010, QCD and high energy interactions}, 1006.2750~[hep-ph].




\vspace*{-0.2cm}\bibitem{Acosta:2004yw}
  D.~E.~Acosta {\it et al.}  [CDF Collaboration],
  Phys.\ Rev.\  D {\bf 71} (2005) 032001.



\vspace*{-0.2cm}\bibitem{boyer}
  B. Boyer {\it et al.} [ALICE Collaboration],
Talk at RQW 2010, Oct. 25-28 2010, Nantes, France
[\href{http://indico.cern.ch/getFile.py/access?contribId=3&sessionId=0&resId=0&materialId=slides&confId=93174}{slides}]

\vspace*{-0.2cm}\bibitem{Ma:2010vd}
  Y.~Q.~Ma, K.~Wang and K.~T.~Chao,
  arXiv:1002.3987 [hep-ph].




\vspace*{-0.2cm}\bibitem{Abe:1997yz}
  F.~Abe {\it et al.}  [CDF Collaboration],
  Phys.\ Rev.\ Lett.\  {\bf 79} (1997) 578.




\vspace*{-0.2cm}\bibitem{Bedjidian:2004gd}
  M.~Bedjidian {\it et al.},
  CERN-2004-009-C, arXiv:hep-ph/0311048. R. Vogt, private communication.




\vspace*{-0.2cm}\bibitem{Khoze:2004eu}
  V.~A.~Khoze, \etal~
  Eur.\ Phys.\ J.\ C {\bf 39} (2005) 163.



\vspace*{-0.2cm}\bibitem{Abelev:2010am}
  B.~I.~Abelev {\it et al.} [ STAR Collaboration ],
  Phys.\ Rev.\  {\bf D82 } (2010)  012004.

\vspace*{-0.2cm}\bibitem{Acosta:2001gv}
  D.~E.~Acosta {\it et al.}  [CDF Collaboration],
  Phys.\ Rev.\ Lett.\  {\bf 88} (2002) 161802.



\vspace*{-0.2cm}\bibitem{CMS-Upsilon}
  CMS~Collaboration, {\it PAS} {\bf BPH-10-003} (2010).


\vspace*{-0.2cm}\bibitem{Affolder:1999wm}
  A.~Affolder {\it et al.}  [CDF Collaboration],
  Phys.\ Rev.\ Lett.\  {\bf 84} (2000) 2094



\vspace*{-0.2cm}\bibitem{Collaboration:2010yr}
  CMS~Collaboration,
  arXiv:1011.4193 [hep-ex].

\vspace*{-0.2cm}\bibitem{ATLAS-JPsi}
  ATLAS~Collaboration,  {\bf ATLAS-CONF-2010-062} (2010).

\vspace*{-0.2cm}\bibitem{LHCb-JPsi}
  LHCb~Collaboration,  {\bf LHCb-CONF-2010-010} (2010).

\vspace*{-0.2cm}\bibitem{Brambilla:2010cs}
  N.~Brambilla {\it et al.}, to appear in Eur.\ Phys.\ J.\  C,
   [arXiv:1010.5827 [hep-ph]].


\end{thebibliography}
\end{document}